\documentclass[11pt]{revtex4}
\usepackage{mathtools}
\usepackage{hyperref}
\usepackage{pdflscape}
\usepackage{graphics}
\usepackage{subfigure}
\usepackage{epsfig}
\usepackage{amsmath,amsfonts}
\usepackage{amssymb}
\usepackage{color}
\usepackage{ulem}
\usepackage{amsmath}
\allowdisplaybreaks

\begin{document}
\title{Gravitational Wave Detection Based on Gravitomagnetic Effects}

\author{Yu-Qi Dong$^{a,b}$}
\email{dongyq2023@lzu.edu.cn}

\author{Zhoujian Cao$^{c}$}
\email{zjcao@amt.ac.cn}

\author{Yu-Xiao Liu$^{a,b}$}
\email{liuyx@lzu.edu.cn (corresponding author)}

\affiliation{$^{a}$ Lanzhou Center for Theoretical Physics, Key Laboratory of Theoretical Physics  of Gansu Province, Key Laboratory of Quantum Theory and Applications of MoE, Gansu Provincial Research Center for Basic Disciplines of Quantum Physics, Lanzhou University, Lanzhou 730000, China\\
	$^{b}$ Institute of Theoretical Physics \& Research Center of Gravitation, Lanzhou University, Lanzhou 730000, China\\
	$^{c}$ Institute for Frontiers in Astronomy and Astrophysics, School of Physics and Astronomy, Beijing Normal University, Beijing, 100875, China\\}

\begin{abstract}
\textbf{Abstract:} In this paper, we explore the feasibility of detecting gravitomagnetic effects generated by gravitational waves,  by monitoring the relative orientation of the angular momentum vectors of test particles. We analyze the response of the relative angular momentum direction to all six polarization modes of gravitational waves and estimate the magnitude of its variation during gravitational wave events.  Our findings indicate that when test particles possess magnetic moments, applying an external magnetic field of appropriate strength can induce resonant precession of the angular momentum direction under the influence of gravitational waves. This resonance may significantly amplify the gravitational wave signal, potentially enabling its detection with future gyroscope-based detectors. Such detectors would complement existing gravitational wave observatories that rely on gravitoelectric effects.

\end{abstract}
	
\maketitle

\section{Introduction}
\label{sec: intro} 

The successful detection of gravitational waves marks the beginning of the era of gravitational wave astronomy for humanity \cite{Abbott1}. Now, in addition to electromagnetic signals, we are able to detect gravitational waves emitted by various sources in the universe. This provides an entirely new channel of information from the universe and deepens our understanding of diverse astrophysical phenomena. As a result, gravitational wave detection has garnered significant attention, opening a new window into astrophysics and cosmology. Furthermore, because various modified gravity theories predict different fundamental properties of gravitational waves—such as their radiation characteristics, polarization modes, and dispersion relations—gravitational wave detection has emerged as a powerful means to test and constrain these theories.

In the language of gravitoelectromagnetic theory \cite{Bahram Mashhoon,XiaoKai He}, for a set of observers with four-velocity $u^{\mu}$ in a given reference frame, the gravitoelectric field $E_{\mu\nu}$ and gravitomagnetic field $B_{\mu\nu}$ can be respectively defined as follows \cite{XiaoKai He}:
\begin{eqnarray}
	\label{definition of the gravitoelectric field and gravitomagnetic field}
	E_{\mu\nu} &\coloneqq& R_{\mu\lambda\nu\rho}u^{\lambda}u^{\rho},
	\nonumber \\
	B_{\mu\nu} &\coloneqq& -\frac{1}{2}\epsilon_{\mu\lambda\rho\sigma}R^{\rho\sigma}_{~~\nu\gamma}u^{\lambda}u^{\gamma}.
\end{eqnarray}	
Here,  $R_{\mu\nu\lambda\rho}$ is the Riemann curvature tensor, and $\epsilon_{\mu\nu\lambda\rho}\coloneqq \sqrt{-g}~e_{\mu\nu\lambda\rho}$ ($e_{0123}=-1$) is the covariant four-dimensional totally antisymmetric tensor in curved spacetime. The detection principles of current gravitational wave detectors are primarily based on the geodesic deviation equation of test particles. Since only the gravitoelectric field $E_{\mu\nu}$ appears in this equation, current detectors exclusively measure the gravitoelectric effects of gravitational waves \cite{J. Aasi,F. Acernese,T. Akutsu,M. Punturo,D. Reitze,P. Amaro-Seoane,Z. Luo,Jun Luo}. Therefore, a natural question arises: can we construct a detector to measure the gravitomagnetic effects of gravitational waves, namely, to detect the $B_{\mu\nu}$ field? The gravitoelectric and gravitomagnetic fields correspond to different components of the Riemann curvature tensor. As a result, detecting the gravitomagnetic effects of gravitational waves can provide additional information about astrophysical sources and the early universe. Moreover, since the physical quantities probed differ from those measured by existing gravitoelectric-based detectors, gravitomagnetic-based gravitational waves detectors could serve as a valuable complement. Combined observations from both types of detectors would enable more comprehensive and stringent tests of general relativity and modified gravity theories.

Due to the equivalence principle, the effects of gravitational waves cannot be detected using a single test particle alone. The relative displacement between two test particles is governed by the gravitoelectric field, while the gravitomagnetic field controls the relative motion of their angular momenta \cite{XiaoKai He,A. Papapetrou,Donato Bini}. This suggests that, in principle, a gravitomagnetic gravitational wave detector can be designed to measure the relative changes in the angular momentum directions of two test particles. The aim of this paper is to explore the feasibility of such a detection scheme.

The paper is organized as follows. In Sec. \ref{sec: 2}, we derive the equation describing the relative motion of test particles' angular momenta. In Sec. \ref{sec: 3}, we investigate the response of the relative angular momentum direction to different polarization modes of gravitational waves. In Sec. \ref{sec: 4}, we estimate the order-of-magnitude of the signal induced in the relative orientation of angular momentum vectors by gravitational wave events. We further uncover a potential resonance mechanism that can significantly amplify the gravitational wave signal, bringing them within reach of future detectors. Section \ref{sec: 5} presents the conclusion.

Unless otherwise specified, we use natural units where $c=G=1$ and adopt the metric signature $(-,+,+,+)$. Greek indices $(\mu,\nu,\lambda,\rho)$ run over four-dimensional spacetime coordinates $(0,1,2,3)$, while Latin indices $(i,j,k,l)$ denote the three spatial directions $(+x,+y,+z)$ and take values from $(1,2,3)$. In the linearized theory, all indices are raised and lowered using the Minkowski metric $\eta_{\mu\nu}$.

\section{Relative Motion of Angular Momentum under Gravitational Waves}
\label{sec: 2}

In this section, we derive the equation of relative motion between the angular momenta of two test particles. Before doing so, we first need to determine the equation of motion for the angular momentum of a single test particle. In fact, the test particle referred to here is an extended body with a finite size, and thus corresponds to the distribution of the energy-momentum tensor 
$T_{\mu\nu}$ within a given small region of space. However, the size of this body is required to be much smaller than the wavelength of the gravitational wave, so that its spatial extent can be neglected when considering the gravitational wave effects. This is why we refer to such an extended body as a test particle. In addition, we require that the velocity of every part of the body is much smaller than the speed of light.

The worldline of such an extended body in spacetime can be represented as a very thin world tube. We consider a coordinate system with spacetime coordinates denoted by $x^{\mu}$. Mathematically, any continuous curve $X^{\mu}(t)$ within this tube can be selected as the  trajectory of the test particle, with $X^{0}=t$ being the time coordinate. Evidently, this choice of trajectory is not unique. With the concept of the trajectory established, we can then define the angular momentum tensor
\begin{eqnarray}
	\label{definition of the angular momentum tensor}
	S^{\mu\nu}(t)
	\coloneqq
	\int d^{3}x \sqrt{-g} \left[\delta x^{\mu}T^{\nu0}(t,x^{i})-\delta x^{\nu}T^{\mu0}(t,x^{i}) \right],
\end{eqnarray}	
where $\delta x^{\mu}\coloneqq x^{\mu}-X^{\mu}(t)$ and the integration is performed over the equal-time hypersurface $x^{0}=t$. It can be shown that $S^{\mu\nu}$ is covariant \cite{A. Papapetrou}. 

It is evident that the definition of $S^{\mu\nu}$ depends on the choice of trajectory $X^{\mu}$, introducing a certain arbitrariness. To establish the connection between $S^{\mu\nu}, X^{\mu}$, with the physically meaningful intrinsic angular momentum and the actual trajectory of the test particle, we need to select a trajectory $X^{\mu}$ that carries clearer physical significance. Such a trajectory can be defined by imposing the condition \cite{C. Moller,K.P. Tod}
\begin{eqnarray}
	\label{PS=0}
	P_{\mu}S^{\mu\nu}=0.
\end{eqnarray}	
Under our assumptions for the test particle, the above $P^{\mu}$ can be approximately defined as \cite{A. Papapetrou}
\begin{eqnarray}
	\label{definition of P}
	P^{\mu}\coloneqq\frac{1}{u^{0}}\int d^{3}x \sqrt{-g} T^{\mu0},
\end{eqnarray}	
where $u^{\mu}\coloneqq dX^{\mu}/d\tau$ with $\tau$ being the proper time of the particle. In the case of gravitational waves under consideration (that is, a weak gravitational perturbation imposed on a flat spacetime background), the trajectory $X^{\mu}$ determined by condition (\ref{PS=0}) corresponds to the center-of-mass coordinate. Therefore, at this point, $X^{\mu}$, $u^{\mu}$, $P^{\mu}$ and $S^{\mu\nu}$ correspond respectively to the actual trajectory, four-velocity, four-momentum and intrinsic angular momentum tensor of the test particle. Furthermore, we can also define a more intuitive angular momentum vector \cite{K.P. Tod}
\begin{eqnarray}
	\label{definition of angular momentum vector}
	S^{\mu}\coloneqq \frac{1}{2}\epsilon^{\mu\nu\lambda\rho}u_{\rho}S_{\nu\lambda}.
\end{eqnarray}	

Applying the condition $\nabla_{\mu}T^{\mu\nu}=0$ along with the assumptions about the test particle, one can derive that the four-velocity and angular momentum of the test particle satisfy (neglect higher-order terms) \cite{A. Papapetrou,XiaoKai He,K.P. Tod}
\begin{eqnarray}
	\label{geodesic equation}
	\frac{Du^{\mu}}{D\tau}=0,
	\\
	\label{angular momentum equation}
	\frac{DS^{\mu}}{D\tau}=0.
\end{eqnarray}	
It can be seen that the particle’s trajectory is a geodesic, and its angular momentum is parallel transported along the geodesic.

We can now derive the equation for the relative motion between the angular momenta of two nearby test particles. Consider two nearby test particles, $A$ and $B$, with positions $X^{\mu}(\tau)$ and $X^{\mu}(\tau)+\eta^{\mu}(\tau)$, respectively, where $\eta^{\mu}$ represents the small displacement between them. Their angular momenta are denoted by $S_{A}^{\mu}$ and $S_{B}^{\mu}$, respectively. According to Eq. (\ref{angular momentum equation}), the angular momenta of particles $A$ and $B$ satisfy the equations
\begin{eqnarray}
	\label{angular momentum equation A}
	\frac{dS_{A}^{\mu}}{d\tau}+\Gamma^{\mu}_{\nu\rho}\left(X^{\mu}\right)\frac{dX^{\nu}}{d\tau}S_{A}^{\rho}&=&0,
	\\
	\label{angular momentum equation B}
	\frac{dS_{B}^{\mu}}{d\tau}+\Gamma^{\mu}_{\nu\rho}\left(X^{\mu}+\eta^{\mu}\right)\frac{d\left(X^{\nu}+\eta^{\mu}\right)}{d\tau}S_{B}^{\rho}&=&0.
\end{eqnarray}	
We choose the proper detector frame, where the trajectory of particle $A$ is $X^{\mu}=(\tau,0,0,0)$, and $g_{\mu\nu}\left(X^{\mu}\right)=\eta_{\mu\nu}$, $\Gamma^{\lambda}_{\mu\nu}\left(X^{\mu}\right)=0$ \cite{MTW,Michele Maggiore}. In this reference frame, Eq. (\ref{angular momentum equation A}) reduces to ${dS_{A}^{\mu}}/{dt}=0$, indicating that the angular momentum of particle $A$ is constant in time. Consequently, the study of the relative change in angular momentum between the two test particles reduces to examining the evolution of particle $B$’s angular momentum $S_{B}^{\mu}$. That is to say, we only need to focus on Eq. (\ref{angular momentum equation B}), which in the proper detector frame can be approximated as
\begin{eqnarray}
	\label{angular momentum equation B in the proper detector frame}
	\frac{dS_{B}^{\mu}}{dt}
	=
	-R^{\mu}_{~\nu\lambda0}\eta^{\lambda}S_{B}^{\nu}.
\end{eqnarray}	
What we actually measure are the spatial components of $S_{B}^{\mu}$. Considering condition (\ref{PS=0}), we can further approximate 
\begin{eqnarray}
	\label{spatial components angular momentum equation B in the proper detector frame}
	\frac{dS_{B}^{i}}{dt}
	=
	-R^{i}_{~jk0}\eta^{k}S_{B}^{j}.
\end{eqnarray}	
Here and below, the Kronecker delta $\delta_{ij}$ and $\delta^{ij}$ are used to lower and raise the three-dimensional space indices, respectively. According to Eq. (\ref{definition of the gravitoelectric field and gravitomagnetic field}), Eq. (\ref{spatial components angular momentum equation B in the proper detector frame}) can equivalently be written as
\begin{eqnarray}
	\label{equation of relative motion}
	\frac{d\overrightarrow{S_{B}}}{dt}
	= \overrightarrow{S_{B}} \times
	\overrightarrow{B_{G}}.
\end{eqnarray}	
Here, $B_{G}^{i}\coloneqq B^{ij}\eta_{j}$. It is evident that the relative motion between the angular momenta of the test particles is influenced by the gravitomagnetic field. In the presence of a constant gravitomagnetic field, $S_{B}^{i}$ undergoes Larmor precession \cite{XiaoKai He}.

\section{Gravitomagnetic Fields and Polarization Modes of Gravitational Waves}
\label{sec: 3}

In general relativity, gravitational waves have only two independent polarization modes: the $+$ mode and the $\times$ mode. However, modified gravity theories often predict additional polarization modes. Thus, the detection of gravitational wave polarizations plays a crucial role in testing and constraining these theories. In fact, it can be shown that under the assumption that matter fields couple only to the metric, gravitational waves in four-dimensional spacetime can have at most six independent polarization modes \cite{Eardley}. This section aims to analyze how the angular momenta of test particles exhibit relative motion under the six polarization modes, thereby laying a theoretical foundation for future gravitational wave detectors that utilize gravitomagnetic effects to probe these polarization modes. Essentially, this requires examining how each of the six gravitational wave polarization modes contributes to the gravitomagnetic field $B_{ij}$.

We now employ gauge invariants to define the six polarization modes of gravitational waves. Within the weak-field approximation, this definition is coordinate-independent, providing the advantage of a more direct connection to observable quantities. The metric perturbation $h_{\mu\nu}\coloneqq g_{\mu\nu}-\eta_{\mu\nu}$ on flat spacetime can be uniquely decomposed into \cite{James. Bardeen,Evans} 
\begin{eqnarray}
	\label{decompose perturbations}
	h_{00}&=&h_{00}, \nonumber \\
	h_{0i}&=&\partial_{i}\gamma+\beta_{i},  \\
	h_{ij}&=&h^{TT}_{ij}+\partial_{i}\epsilon_{j}+\partial_{j}\epsilon_{i}
	+\frac{1}{3}\delta_{ij}H+\left(\partial_{i}\partial_{j}-\frac{1}{3}\delta_{ij}\Delta\right)\zeta\nonumber,
\end{eqnarray}
where $\beta^{i}$ and $\epsilon^{i}$ are transverse, and $h^{TT}_{ij}$ is transverse and traceless:
\begin{eqnarray}
	&\partial_{i}\beta^{i}=\partial_{i}\epsilon^{i}=0,  \\
	&\delta^{ij}h^{TT}_{ij}=0,~\partial^{i}h^{TT}_{ij}=0.
\end{eqnarray}
From this, we can construct the following gauge invariants: one spatial tensor
\begin{eqnarray}
	\label{tensor gauge invariant}
	h^{TT}_{ij},
\end{eqnarray}
one spatial vector
\begin{eqnarray}
	\label{vector gauge invariant}
	\begin{array}{l}
		\Xi_{i}\coloneqq\beta_{i}-\partial_{0}\epsilon_{i},
	\end{array}	
\end{eqnarray}
and two spatial scalars
\begin{eqnarray}	
	\begin{array}{l}
		\phi \,\coloneqq -\frac{1}{2}h_{00}+\partial_{0}\gamma-\frac{1}{2}\partial_{0}\partial_{0}\zeta,  \\
		\Theta \coloneqq \frac{1}{3}\left(H-\Delta\zeta\right).
	\end{array} \label{scalar gauge invariant}
\end{eqnarray}
They remain invariant under the gauge transformation $h_{\mu\nu} \rightarrow	h_{\mu\nu}-\partial_{\mu}\xi_{\nu}-\partial_{\nu}\xi_{\mu}$ ($\xi^{\mu}$ is an arbitrary function). When analyzing a plane gravitational wave propagating on a flat background, we can, without loss of generality, choose the propagation direction to be along the $+z$-axis. In this setup, the six polarization modes of the gravitational wave can be defined as \cite{Eardley,Yu-Qi Dong}
\begin{eqnarray}
	\label{P1-P6 gauge invariant}
	\begin{array}{l}
		P_{1}\coloneqq\partial_{3}\partial_{3}\phi-\frac{1}{2}\partial_{0}\partial_{0}\Theta, \quad
		P_{2}\coloneqq\frac{1}{2}\partial_{0}\partial_{3}\Xi_{1},\\
		P_{3}\coloneqq\frac{1}{2}\partial_{0}\partial_{3}\Xi_{2},  \quad \quad\quad\quad\,\,
		P_{4}\coloneqq-\frac{1}{2}\partial_{0}\partial_{0}h^{TT}_{11}, \\
		P_{5}\coloneqq-\frac{1}{2}\partial_{0}\partial_{0}h^{TT}_{12}, \quad\quad~~~
		P_{6}\coloneqq-\frac{1}{2}\partial_{0}\partial_{0}\Theta.
	\end{array}
\end{eqnarray}
Here, $P_{4}$ and $P_{5}$ represent the well-known $+$ and $\times$ modes, which are classified as tensor modes. $P_{2}$ and $P_{3}$ are known as the vector-$x$ and vector-$y$ modes, respectively, and are classified as vector modes. $P_{1}$ and $P_{6}$ correspond to the longitudinal and breathing modes, respectively, and are classified as scalar modes.

Since the Riemann curvature tensor $R_{\mu\nu\lambda\rho}$ is gauge invariant, it follows from Eq. (\ref{definition of the gravitoelectric field and gravitomagnetic field}) that the gravitomagnetic field is also gauge-invariant. Furthermore, by applying Eqs. (\ref{tensor gauge invariant}), (\ref{vector gauge invariant}), and (\ref{scalar gauge invariant}), the spatial components of the gravitomagnetic field can be written as
\begin{eqnarray}
	\label{Bij}
	B_{ij}=\frac{1}{2}\begin{pmatrix}
		\partial_{0}\partial_{3}h^{TT}_{12}~ & 
		-\partial_{0}\partial_{3}h^{TT}_{11}+\partial_{0}\partial_{3}\Theta~ & 
		-\partial_{3}\partial_{3}\Xi_{2}\\
		-\partial_{0}\partial_{3}h^{TT}_{11}-\partial_{0}\partial_{3}\Theta~    &
	    -\partial_{0}\partial_{3}h^{TT}_{12}~  &
	    \partial_{3}\partial_{3}\Xi_{1}~\\
		0       &  0   &   0
	\end{pmatrix}.
\end{eqnarray}
It is evident that the $+$ mode, caused by $h^{TT}_{11}$, contributes equally to the $B_{12}$ and $B_{21}$ components. Similarly, the $\times$ mode contributes to $B_{11}$ and $B_{22}$, the vector-$x$ mode contributes to $B_{23}$, and the vector-$y$ mode contributes to $B_{13}$. For scalar modes, the situation is more intricate. This is because only $\Theta$ contributes to $B_{ij}$, while $\Theta$ simultaneously excites both the breathing and longitudinal modes. Given that the longitudinal mode depends on both $\Theta$ and $\phi$, while the breathing mode depends only on $\Theta$, this implies that detection via the gravitomagnetic effect can probe the breathing mode but not the longitudinal mode. Detecting the latter requires gravitational wave detectors based on gravitoelectric effects. Using Eq. (\ref{Bij}), the relative motion between the angular momenta of the test particles under various polarization modes can be easily inferred from Eq. (\ref{equation of relative motion}).

Finally, combined detection using gravitoelectric and gravitomagnetic effect-based detectors can effectively determine the speed of gravitational waves. Assuming the gravitational wave propagates at speed $v$, the perturbation $h_{\mu\nu}$ depends solely on $u\coloneqq t-z/v$. In this case, the gravitomagnetic field satisfies
\begin{eqnarray}
	\label{Bij with v}
	B_{ij}=\frac{1}{2}\begin{pmatrix}
		-{\ddot{h}^{TT}_{12}}/{v}~ & 
		{\ddot{h}^{TT}_{11}}/{v}-{\ddot\Theta}/{v}~ & 
		-\ddot\Xi_{2}/v^{2}\\
		\ddot{h}^{TT}_{11}/v+\ddot\Theta/v~    &
		\ddot{h}^{TT}_{12}/v~  &
		\ddot\Xi_{1}/v^{2}~\\
		0       &  0   &   0
	\end{pmatrix},
\end{eqnarray}
while the gravitoelectric field satisfies \cite{Yu-Qi Dong}
\begin{eqnarray}
	\label{Eij with v}
	E_{ij}=\frac{1}{2}\begin{pmatrix}
		-\ddot{h}^{TT}_{11}-\ddot\Theta~ & 
		-\ddot{h}^{TT}_{12}~ & 
		-\ddot\Xi_{1}/v\\
		-\ddot{h}^{TT}_{12}~    &
		\ddot{h}^{TT}_{11}-\ddot\Theta~  &
		-\ddot\Xi_{2}/v~\\
		-\ddot\Xi_{1}/v~      &
		  -\ddot\Xi_{2}/v~   &
		  2\ddot{\phi}/v^{2}-\ddot\Theta
	\end{pmatrix}.
\end{eqnarray}
Here and throughout the following text, a dot above a variable denotes its time derivative. It is clear that once detectors measure both the gravitoelectric and gravitomagnetic fields, the speed of gravitational waves can be directly determined from the ratio of the gravitoelectric to gravitomagnetic fields for the corresponding polarization mode. For instance, for the vector-$x$ mode, the speed $v=-E_{13}/B_{23}$.

\section{Resonant Amplification of Gravitational Wave Signals}
\label{sec: 4}

\begin{figure*}[htbp]
	\makebox[\textwidth][c]{\includegraphics[width=0.6\textwidth]{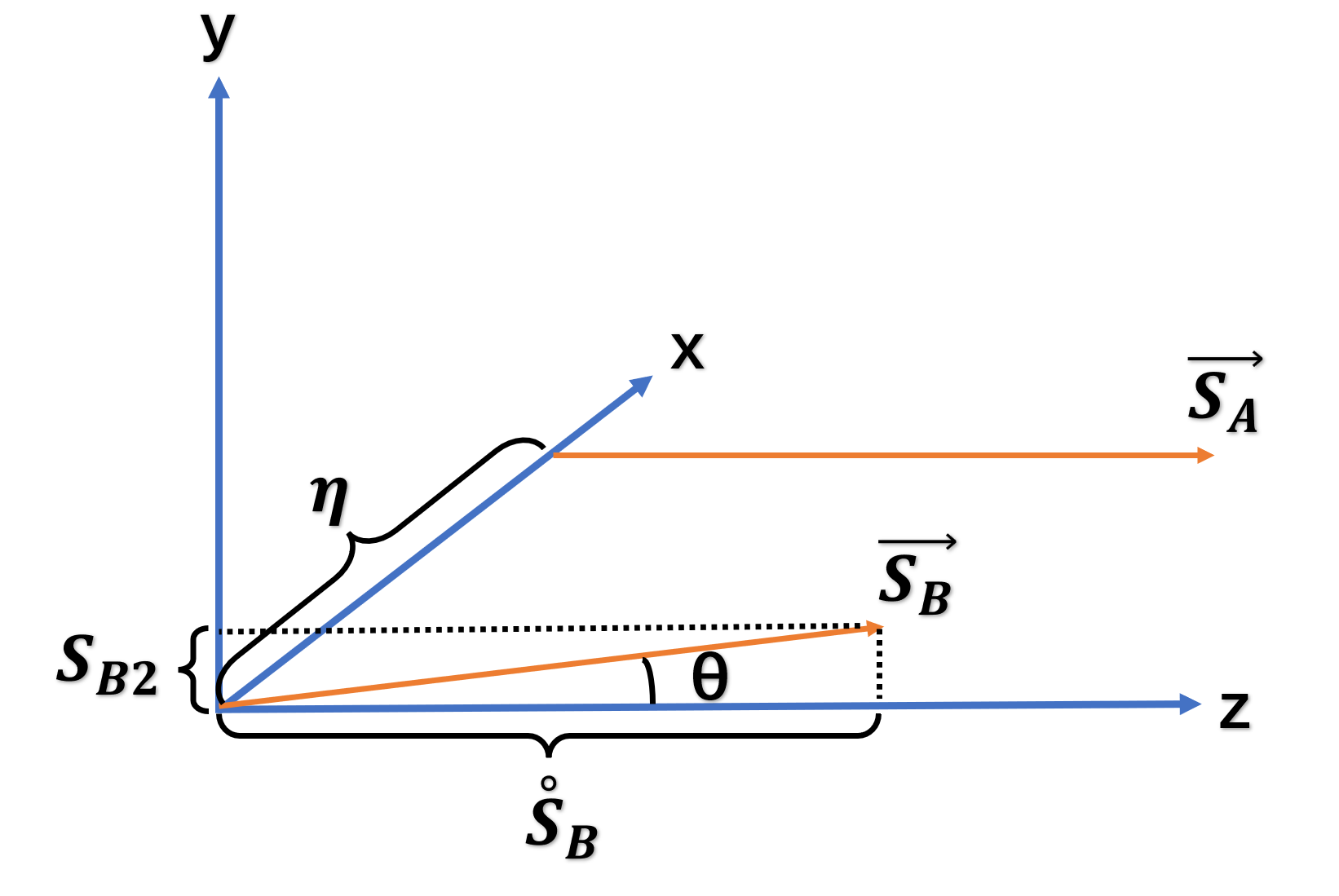}}
	\caption{Relative motion of the angular momenta of two test particles under a gravitational wave in the proper detector frame.}
	\label{fig: 1}
\end{figure*}

We now estimate the order of magnitude of the relative change in angular momentum direction induced by a gravitational wave. As an example, we consider a $\times$ mode gravitational wave propagating in the $+z$ direction. The initial direction of $\overrightarrow{S_{B}}$
is aligned with $+z$: $\overrightarrow{\mathring{S}_{B}}=(0,0,\mathring{S}_{B})$, while the relative displacement between the two test particles is along the $+x$ axis: $\eta^{i}=(\eta,0,0)$. According to Eqs. (\ref{equation of relative motion}) and (\ref{Bij}), $\overrightarrow{S_{B}}$ undergoes a small periodic angular oscillation in the $y$-direction, with angular velocity given by 
\begin{eqnarray}
	\label{angular velocity}
	\dot{\theta}=\frac{\dot{S}_{B2}}{\mathring{S}_{B}}=-\frac{1}{2}\frac{\ddot{h}^{TT}_{12}\eta}{v}=-2\pi^{2}f h^{TT}_{12}\frac{\eta}{\lambda},
\end{eqnarray}
as shown in Fig. \ref{fig: 1}. Here, $f$ and $\lambda$ denote the frequency and wavelength of the gravitational wave, respectively.
To ensure the validity of Eq. (\ref{equation of relative motion}), $\eta$ should be much smaller than $\lambda$. Typically, one can take $\eta/\lambda \sim 10^{-2}$. For gravitational wave events detectable by ground-based detectors, $h^{TT}_{12}\sim 10^{-22}$ and $f \sim 10^{2}~Hz$ \cite{Abbott2}. Such events induce $\dot{\theta} \sim 10^{-21}~rad/s$. For gravitational wave events detectable by space-based detectors, the typical values are $h^{TT}_{12}\sim 10^{-19}$, $f \sim 10^{-2}~Hz$ \cite{P. Amaro-Seoane2}, and $\dot{\theta} \sim 10^{-22}~rad/s$. This is far beyond the current sensitivity of gyroscopes. For instance, with an integration time of approximately $2\times 10^{5} s$, the upper limit of the angular velocity measurement precision for the GINGERINO active-ring laser gyroscope is close to $2 \times 10^{-15}~rad/s$ \cite{A. D. V. Di Virgilio}. This is at least six orders of magnitude greater than the gravitational wave signal. The challenge now is to find physical mechanisms capable of amplifying gravitational wave signals, enabling their detection by current or next-generation high-precision gyroscopes.

Equation (\ref{equation of relative motion}) bears a formal resemblance to the equation describing the motion of a magnetic moment in a magnetic field. This analogy naturally inspires the idea of amplifying gravitational wave signals through a mechanism similar to magnetic resonance. Assuming that test particle $B$ has a magnetic moment proportional to its angular momentum $\overrightarrow{S_{B}}$, its angular momentum will also be influenced by an external magnetic field:
\begin{eqnarray}
	\label{equation of relative motion in B}
	\frac{d\overrightarrow{S_{B}}}{dt}
	= \overrightarrow{S_{B}} \times
	\overrightarrow{B_{G}}
	+\gamma \overrightarrow{S_{B}} \times \overrightarrow{B}.
\end{eqnarray}	
Here, $\overrightarrow{B}$ is the external magnetic field, and $\gamma$ denotes the parameter representing the ratio of the test particle’s magnetic moment to its angular momentum. Thus, $\overrightarrow{S_{B}}$ is simultaneously influenced by both the gravitomagnetic field and the external magnetic field. In the absence of gravitational waves, applying a constant magnetic field $\overrightarrow{B}$ causes $\overrightarrow{S_{B}}$ to undergo Larmor precession around the $\overrightarrow{B}$-axis at a fixed frequency. If a gravitational wave arrives, it can be shown that when its frequency matches the Larmor precession frequency of $\overrightarrow{S_{B}}$, resonance occurs, significantly amplifying $\overrightarrow{S_{B}}$'s response to the gravitational wave.

To further illustrate this, we revisit the example at the beginning of this section, now with the addition of a constant external magnetic field $\overrightarrow{B}=(0,0,B)$ aligned along the $+z$ direction. Thus, according to Eq. (\ref{equation of relative motion in B}), the various components of $\overrightarrow{S_{B}}$ satisfy the following equations:
\begin{eqnarray}
	\label{spatial components SB1}
	\dot{S}_{B1}&=&\gamma B S_{B2},
	\\
	\label{spatial components SB2}
	\dot{S}_{B2}&=&-\gamma B S_{B1}
	-2\pi^{2}f \mathring{S}_{B} h^{TT}_{12}\frac{\eta}{\lambda},
	\\
	\label{spatial components SB3}
	\dot{S}_{B3}&=&0.
\end{eqnarray}
It can be seen that under the influence of the gravitational wave, the angular momentum undergoes variations in the $x$ and $y$ directions. To better reveal the resonance effect, we take the time derivative of Eq. (\ref{spatial components SB2}) and combine it with Eq. (\ref{spatial components SB1}). As a result, $S_{B2}$ satisfies the following equation:
\begin{eqnarray}
    \label{spatial components SB2 motion}
	\ddot{S}_{B2}+\omega_{0}^{2}S_{B2}=
	-2\pi^{2}f \mathring{S}_{B} \dot{h}^{TT}_{12}\frac{\eta}{\lambda},
\end{eqnarray}
where $\omega_{0} \coloneqq \gamma B$. This equation describes a driven harmonic oscillator. Resonance occurs when the frequency of the driving force is equal to the oscillator’s natural frequency $\omega_{0}$. When $h^{TT}_{12}\propto e^{i \omega t}$, a particular solution for $S_{B2}$ is given by
\begin{eqnarray}
	\label{particular solution of SB2}
	\frac{S_{B2}}{\mathring{S}_{B}}=\frac{i \pi \omega^{2}}{\omega^{2}-\omega_{0}^{2}}\frac{\eta}{\lambda} h^{TT}_{12}.
\end{eqnarray}
Comparing with Eq. (\ref{angular velocity}), the resonance effect enhances $\dot{S}_{B2}/\mathring{S}_{B}$ by a factor of by a factor of $\left| \omega^{2}/\left(\omega^{2}-\omega_{0}^{2}\right) \right|$. In theory, if $\omega_{0}$ approaches the gravitational wave frequency $\omega$ arbitrarily closely, the signal can, in principle, be infinitely amplified. 

Test particles that can exhibit the resonance effect may be fabricated from magnetic materials with a permanent magnetic moment. For example, consider a rotor consisting of an NdFeB sphere with radius $R$, rotating at angular velocity $\bar{\omega}$, where the magnetization direction is aligned with the rotation axis and the magnetization is $M$. In this way, we have $\gamma=\frac{5M}{2\rho\bar{\omega}R^{2}}$, where $\rho$ is the material density. For NdFeB, $M\sim 10^{6}~A/m$, and $\rho \sim 7.5 \times 10^{3}~kg/m^{3}$. Therefore, in the case of $R \sim 10^{-2}~m$ and $\bar{\omega} \sim 10^{3}~rad/s$, $\gamma \sim 10^{3} ~rad/s/T$. To match $\omega_{0}$ with the characteristic frequencies of typical gravitational wave events, the required external magnetic field ranges from a few gauss to several tesla, which can be reliably generated in laboratory conditions.
In practical detection, the factor $\left| \omega^{2}/\left(\omega^{2}-\omega_{0}^{2}\right) \right|$ is limited by the precision with which we can apply the precession frequency $\omega_{0}$ to the object. These limitations introduce a relative error $\delta \coloneqq \left(\omega_{0}-\omega\right)/\omega \ll 1$. Therefore, the amplification factor is $\left| 1/\left(2\delta\right) \right|$. Clearly, effective signal amplification requires fabricating rotors with extremely low relative error in $\gamma$ and applying highly stable magnetic fields. Currently, the relative precision of magnetic fields achievable in laboratories can reach $10^{-6}$ \cite{Mateusz Borkowski}. The parameter $\gamma$ of the material can be precisely determined by measuring its precession frequency in an external magnetic field. Due to the exceptional timing precision of current atomic clocks, $\gamma$ can theoretically be measured with a relative accuracy matching that of the magnetic field. Furthermore, the stability of $\gamma$ is also related to the stability of the material's rotational speed and its magnetic moment \cite{C Adambukulam}, which can be effectively maintained under vacuum and low-temperature conditions. Consequently, future technological advancements may enable amplification factors reaching $10^{6}$ or higher, thereby facilitating the detection of gravitational wave signals using instruments such as superconducting quantum interference device (SQUID) gyroscopes. {Finally, realizing the amplification effect requires prior knowledge of the gravitational wave frequency, so that the magnetic-field-induced precession can be tuned into resonance with it. Although the exact frequency of transient events cannot be predetermined, for long-lived continuous gravitational wave sources with stable frequencies, the wave frequency can be accurately determined from existing detectors or astrophysical observations. Once the frequency is known, the external magnetic field can be adjusted accordingly to achieve the resonance condition. Therefore, detectors based on this resonance mechanism may be particularly suited for continuous gravitational waves with stable frequencies.}


\section{Conclusion}
\label{sec: 5}

In this paper, we investigated the feasibility of probing gravitational waves through their gravitomagnetic effects. The gravitoelectric and gravitomagnetic fields correspond to different components of the Riemann curvature tensor. Consequently, detecting the gravitomagnetic effect of gravitational waves can provide information beyond that accessible with current detectors based on the geodesic deviation equation. This additional information can enhance our ability to measure physical quantities, test gravitational theories, and expand gravitational wave astronomy. In contrast to the gravitoelectric effect, which results in relative displacements between test particles, the gravitomagnetic effect of gravitational waves leads to relative motion in their angular momenta. This motion is governed by Eq. (\ref{equation of relative motion}), suggesting that the gravitomagnetic effect of gravitational waves can be probed by measuring the relative precession of angular momenta.

In four-dimensional spacetime, assuming matter fields couple solely to the metric, gravitational waves can exhibit up to six independent polarization modes. Employing gauge invariants, we analyzed the relative motion of angular momenta induced by these six modes. Our analysis shows that detection via the gravitomagnetic effect can identify up to five of the six polarization modes, but cannot ascertain the existence of the longitudinal mode. Moreover, by jointly measuring the gravitoelectric and gravitomagnetic fields, the propagation speed of gravitational waves can, in principle, be directly determined from the ratio between these fields.

We also estimated the order of magnitude of the relative precession angular velocity of angular momenta induced by gravitational wave events. The results indicate that the gravitational wave signals are below the detectable threshold of the most sensitive current gyroscopes, 
rendering them undetectable with existing technology. However, the formal structure of Eq. (\ref{equation of relative motion}) is identical to that of the Larmor precession equation, suggesting that gravitational waves could be amplified by a magnetic-resonance-like mechanism.
This resonance mechanism may be implemented using a rotor fabricated from magnetic material with magnetization aligned along its axis of rotation. The rotor’s angular momentum would couple simultaneously to an external magnetic field and to the gravitational field, thereby obeying Eq. (\ref{equation of relative motion in B}), which has the same form as the equation governing magnetic resonance. Through precision fabrication of the rotor, application of a highly stable magnetic field, and careful maintenance of both the rotational speed and magnetic moment stability, this configuration could enhance the response induced by gravitational waves by several orders of magnitude. 
Such amplification would significantly relax the measurement precision required to detect the gravitomagnetic effect of gravitational waves, thereby making its observation more feasible with future advanced instrumentation. However, employing this resonance effect requires prior knowledge of the target gravitational wave frequency. Since determining the required frequency with sufficient precision typically takes time, detectors based on this resonance amplification mechanism may be particularly well suited for observing long-lived gravitational wave signals with stable frequencies. This study offers only a preliminary estimate of a potential instrument for detecting the gravitomagnetic effect of gravitational waves. A more detailed design and comprehensive feasibility assessment will be addressed in future work.

\section*{Acknowledgments}
This work is supported in part by the National Natural Science Foundation of China (Grants No. 123B2074, No. 12475056,  and No. 12247101), Gansu Province's Top Leading Talent Support Plan, the Fundamental Research Funds for the Central Universities (Grant No. lzujbky-2025-jdzx07), the Natural Science Foundation of Gansu Province (No. 22JR5RA389 and No.25JRRA799), and the `111 Center’ under Grant No. B20063.


\end{document}